\documentclass[aps,prb,superscriptaddress,showpacs,twocolumn]{revtex4-1}

\usepackage{amsmath}
\usepackage{graphicx}

\begin{document}

\title{Determining molecule diffusion coefficients on surfaces from a
  locally fixed probe:\\ On the analysis of signal fluctuations }
 
\author{Susanne Hahne}
\affiliation{Fachbereich Physik, Universit\"at Osnabr\"uck,
Barbarastra{\ss}e 7, 49076 Osnabr\"uck, Germany}

\author{Julian Ikonomov}
\affiliation{Institut f\"ur Physikalische und Theoretische Chemie, 
Universit\"at Bonn, Wegelerstra{\ss}e 12, 53115 Bonn, Germany}

\author{Moritz Sokolowski}
\affiliation{Institut f\"ur Physikalische und Theoretische Chemie, 
Universit\"at Bonn, Wegelerstra{\ss}e 12, 53115 Bonn, Germany}

\author{Philipp Maass}
\email{philipp.maass@uni-osnabrueck.de}
\affiliation{Fachbereich Physik, Universit\"at Osnabr\"uck,
Barbarastra{\ss}e 7, 49076 Osnabr\"uck, Germany}

\date{December 7, 2012}

\begin{abstract}
  Methods of determining surface diffusion coefficients of molecules
  from signal fluctuations of a locally fixed probe are revisited and
  refined.  Particular emphasis is put on the influence of the
  molecule's extent. In addition to the formerly introduced
  autocorrelation method and residence time method, we develop a
  further method based on the distribution of intervals between
  successive peaks in the signal. The theoretical findings are applied
  to STM measurements of copper phthalocyanine (CuPc) molecules on the
  Ag(100) surface.  We discuss advantages and disadvantages of each
  method and suggest a combination to obtain accurate results for
  diffusion coefficients.
\end{abstract}


%68.43.Jk Diffusion of adsorbates, kinetics of coarsening and aggregation
%68.35.Fx Diffusion; interface formation (see also 66.30.-h 
%         Diffusion in solids, for diffusion of adsorbates, see 68.43.Jk)
%82.37.Gk STM and AFM manipulations of a single molecule (for atom
%         manipulation see 37.10.Gh, Pq in atomic and molecular physics; see
%         also 81.16.Ta Atom manipulation in methods of nanofabrication and
%         processing; 87.80.Nj Single-molecule techniques in biological physics)

\maketitle

\section{Introduction}\label{sec:intro}

To describe the growth kinetics of adsorbates on surfaces, knowledge
of diffusion coefficients of atoms and molecules on the surface is of
vital importance. Different experimental techniques can be utilized to
measure corresponding diffusion coefficients.  Two of the first
methods were Field Emission Microscopy\cite{Mueller:1938,Gomer:1990}
and Field Ionization Microscopy.\cite{Mueller/Tsong:1969} Another
technique is the laser induced thermal
desorption,\cite{Viswanathan/etal:1982,George/etal:1985} where first
an area of adparticles is cleared by desorption and then the
successive repopulation measured by desorption. For scanning
tunneling microscopy (STM) several modes of operation have been
described: Taking scans of the surface with the adsorbed particles at
different time instants to identify individual particle
displacements,\cite{Ganz/etal:1992} recording current signals in time
with a locally fixed STM tip,\cite{Gomer:1986} and coupling the tip to
a single adsorbate and tracking its motions.\cite{Swartzentruber:1996}
Different procedures to extract diffusion coefficients from these and
other techniques have been reviewed by Barth in 2000.\cite{Barth:2000}

Analysis of island densities with predictions from rate equations for
submonolayer growth kinetics is a further powerful method,
\cite{Brune:1998,Evans/etal:2006} which allows one also to determine
binding energies between adatoms or of adatoms to small clusters. An
extension of this method to multi-component systems is given in
Refs.~\onlinecite{Einax/etal:2007b,Einax/etal:2009}.

In recent years the self-assembly of organic molecules on surfaces has
attracted much attention in connection with the idea of developing
electronic devices on a molecular level.\cite{Kuehnle:2009} Molecules
in organic surface growth often have high mobilities and sizes large
compared to the lattice constant of the substrate. In this situation
not all techniques are equally well suited. A convenient and powerful
method is the recording of temporal current fluctuations with a fixed
scanning tunneling microscope tip\cite{Lozano/Tringides:1995} or, in
principle, of signal fluctuations from any locally fixed microscopic
or other probe.\cite{Gomer:1990} For example, for molecules on
insulating substrates, the frequency or height fluctuations in an
atomic force microscopy could be used.

In order to extract diffusion coefficients from signal fluctuations of
a probe, a theoretical basis for the analysis of the measurement is
necessary. One approach was developed for the auto-correlation
function (ACF) of an STM signal \cite{Sumetskii/Kornyshev:1993,
  Sumetskii/etal:1994} and applied, on the basis of the corresponding
power spectrum, to oxygen atom diffusion on
Si(111),\cite{Lozano/Tringides:1995} and more recently to hydrogen
atom diffusion on Si(111).\cite{Tringides/Hupalo:2010} Another
approach was followed in Ref.~\onlinecite{Ikonomov/etal:2010} by
analyzing the distribution of peak widths in the signal, which
correspond to residence times of single molecules under an STM tip. We
will refer to it as the residence time distribution (RTD) method in
the following.  For determining absolute values of diffusion
coefficients, a general problem is that the theoretical modeling
involves a commonly not precisely known ``detection function'' with
some finite range. However, when the extension of the diffusing
molecules is larger than the detection range, this problem becomes
essentially irrelevant and the size of the molecule enters as the
relevant length scale.

In this work we first revisit both the ACF and the RTD method.  For
the ACF method we study, as a new aspect, modifications implied by the
finite size of molecules and for the RTD method we give a detailed
account of the functional form of the RTD in different time regimes.
In the RTD, time intervals are considered where the molecule is under
(or very close to) the tip. Corresponding time intervals are also
included in the calculation of the ACF. It can therefore not be
precluded that the interaction of the molecule with the tip is
influencing the results. A possible influence can be evaluated
systematically by changing the bias voltage,\cite{Ikonomov/etal:2010}
or the distance between tip and substrate.  In order to ensure that
the influence is small independent of adjustments, we develop a
further method, which is based on the analysis of the distribution of
peak-to-peak distances in the signal. This method is referred to as
the interpeak time distribution (ITD) method in the following.

All three methods are applied to diffusion measurements of copper
phthalocyanine (CuPc) on the Ag(100) surface previously reported in
Ref.~\onlinecite{Ikonomov/etal:2010}. We discuss advantages and
disadvantages of the three procedures mentioned above and a possible
combination of them, which enables us to improve the quality of
results for the diffusion coefficient.

\section{Auto-correlation function (ACF)}\label{sec:acf}
We consider $N$ molecules diffusing on a flat substrate area $A$ and a
probing tip of an STM placed at a fixed position above the surface.
The lateral tip position is chosen as the origin of an
$x$-$y$-coordinate system on the surface. Each diffusing molecule
$\alpha$ with its center at position
$\mathbf{r}_\alpha(t)=(x_\alpha(t),y_\alpha(t))$ gives a contribution
$i_\alpha(t)=G(\mathbf{r}_\alpha(t))$ to the total tunneling current
$I(t)=\sum_{\alpha=1}^N i_\alpha(t)$, where $G(\mathbf{r})$ is the
``detection function''.  We regard the molecules as non-interacting,
which should be a valid assumption at low concentrations $c=N/A$, or
more precisely on time scales smaller than $l^2/D$, where $l\sim
c^{-1/2}$ is the mean intermolecular distance and $c$ the number
density of molecules. The autocorrelation function $C(t)=\langle
\delta I(0)\delta I(t)\rangle$ of the current fluctuations $\delta
I(t)=I(t)-\langle I\rangle$ can then be written as
\begin{eqnarray}
C(t) &=& N\langle i(0)i(t)\rangle\nonumber\\
     &=& c \int \textrm{d}^2\mathbf{r}_0 
\int \textrm{d}^2\mathbf{r}_1 G(\mathbf{r}_1) 
f(\mathbf{r}_1,t|\mathbf{r}_0) G(\mathbf{r}_0)\,,
\label{eq:K}
\end{eqnarray}
where $f(\mathbf{r}_1,t|\mathbf{r}_0)$ is the two-dimensional
diffusion propagator
\begin{equation}
f(\mathbf{r}_1,t|\mathbf{r}_0) = \frac{1}{4\pi D t}
\exp\left(-\frac{(\mathbf{r}_1-\mathbf{r}_0)^2}{4 D t}\right)\,.
\label{eq:f}
\end{equation}

To provide a quantitative prediction for $C(t)$, knowledge of the
detection function $G(\mathbf{r})$ is necessary. For the STM this
requires a detailed treatment of the tunneling problem, which is a
long-standing problem since the first pioneering work by Tersoff and
Hamann. \cite{Tersoff/Hamann:1983} Based on a simple approach,
Sumetskii and Kornyshev \cite{Sumetskii/Kornyshev:1993} suggested a
Gaussian detection function for small particles (adatoms) as a result
of the tunneling barrier. For extended molecules, the lateral
structure of the electronic charge density needs to be taken into
account. In fact, without referring to first-principle calculations
for the specific system under consideration, it is difficult to obtain
accurate expressions for the tunneling current.

While such more detailed treatments are interesting in themselves,
they are not really required for determining the diffusion coefficient
$D$. A more practical approach is to transform $I(t)$ into a
rectangular signal $I_\textrm{rec}(t)=\Theta(I(t)-I_c)$, where
$\Theta(.)$ is the Heaviside jump function ($\Theta(x)=1$ for $x>0$
and zero else), and $I_c$ is a threshold current to exclude background
noise.  If we consider situations, where at most one molecule can be
in the detection region under the tip, each single molecule
contribution $i_\alpha(t)$ to $I_\textrm{rec}(t)$ also transforms to a
rectangular signal $i^\textrm{rec}_\alpha(t)\in\{0,1\}$, and
$I_\textrm{rec}(t)=\sum_\alpha i^\textrm{rec}_\alpha(t)\in\{0,1\}$.
Accordingly, the detection function can be defined as ``indicator
function'' of the molecule
\begin{equation}
G(\mathbf{r})=\left\{\begin{array}{ll}
1 & \textrm{if\ } \mathbf{r}\in\mathcal{M}\\
0 & \textrm{else\,,}\end{array}\right.
\label{eq:g}
\end{equation}
where $\mathcal{M}$ is the set of center positions of the molecule
which give rise to a tunneling current $I>I_c$. To get a description
that is independent of details of the molecule's shape, we assign a
circle to the set $\mathcal{M}$. The radius $R$ of this circle can,
for example, be determined by setting the covered area of the molecule
equal to $\pi R^2$, by the gyration radius of the set $\mathcal{M}$,
or by some other reasonable requirement. The function $G(\mathbf{r})$
from Eq.~(\ref{eq:g}) is then approximated by
$\Theta(R-|\mathbf{r}|)$.

The mutual exclusion under the tip implies that the molecules are no
longer non-interacting as assumed when deriving Eq.~(\ref{eq:K}). This
leads to nonzero cross-correlations $\langle
i_\alpha^\textrm{rec}(t)i_\beta^\textrm{rec}(0)\rangle$,
$\alpha\ne\beta$.  While these cross-correlations could be treated
approximately, we concentrate here on times smaller than the mean
residence time $~\sim \tau_R:=R^2/D$ of a molecule under the tip. In
this regime we can set $\langle
i_\alpha^\textrm{rec}(t)i_\beta^\textrm{rec}(0)\rangle\simeq0$ for
$\alpha\ne\beta$. Accordingly, the self-part $N\langle
i_\alpha^\textrm{rec}(t)i_\alpha^\textrm{rec}(0)\rangle$ appearing in
Eq.~(\ref{eq:K}) gives the correlation function
$C_\textrm{rec}(t)=\langle I_\textrm{rec}(t)I_\textrm{rec}(0)\rangle$
(i.e.\ without subtracting a term $\langle I_\textrm{rec}\rangle^2$).
Evaluation of Eq.~(\ref{eq:K}) then yields
\begin{align}
C_\textrm{rec}(t)&=4\pi c\int_0^R\textrm{d}r_0
r_0\int_0^R\textrm{d}r_1 r_1\label{eq:krec}\\
&\hspace{6em}{}\times\frac{\exp\left(-\frac{r_0^2+r_1^2}{4Dt}\right)}{4Dt}\,
I_0\left(\frac{2r_0r_1}{4Dt}\right)\,,
\nonumber
\end{align}
where $I_0$ is the modified Bessel function of zeroth order.

The knowledge of $C_\textrm{rec}(t)$ for times $t<\tau_R$, as
predicted by Eq.~(\ref{eq:krec}) for single-particle diffusion, is
sufficient to determine $D$.  For $t>\tau_R$, collective properties
and the associated cross-correlations had to be taken into account. In
a rough treatment one could factorize $\langle
i_\alpha^\textrm{rec}(t)i_\beta^\textrm{rec}(0)\rangle\simeq \langle
i_\alpha^\textrm{rec}(t)\rangle\langle
i_\beta^\textrm{rec}(0)\rangle$, which amounts to add $(c\pi R^2)^2$
to the expression in Eq.~(\ref{eq:krec}) for $t\gg \tau_R$.

\begin{figure}[t!]
\includegraphics[height=4cm]{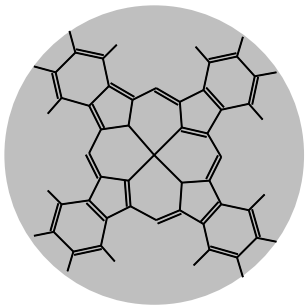}
\caption{Relaxed structure of the CuPc molecule. Notice that the
  terminating bonds are connected to hydrogen atoms. The assigned
  detection area marked in gray is a circle with radius $R=7.6$~\AA.
  It corresponds to the smallest circle that entails all nuclei.}
\label{fig:cupc}
\end{figure}

\begin{figure}[b!]
\includegraphics[width=8cm]{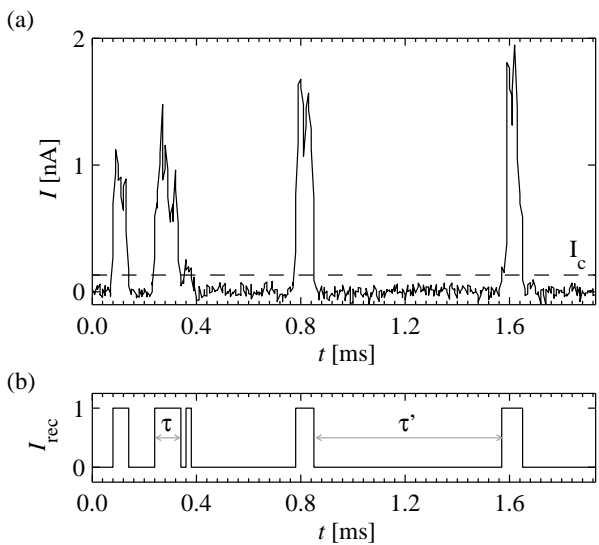}
\caption{(a) Fluctuating STM tunneling current $I(t)$ observed during
  CuPc diffusion on Ag(100) at $T=192$~K ($U_{\textrm{bias}}$=1.7~V,
  $I_{\textrm{setpoint}}$=20~pA). The dashed line marks the threshold
  value $I_c$, above which the distribution of the current values has
  no longer a Gaussian shape, see Fig.~\ref{fig:noiselevel}. (b) The
  rectangular signal $I_\textrm{rec}(t)=\Theta(I(t)-I_c)$ associated
  with $I(t)$.  The widths of the rectangles give the residence times
  $\tau$ and the distances between the rectangles give the interpeak
  times $\tau'$.}
\label{fig:currents}
\end{figure}

\begin{figure}[t!]
\includegraphics[width=8cm]{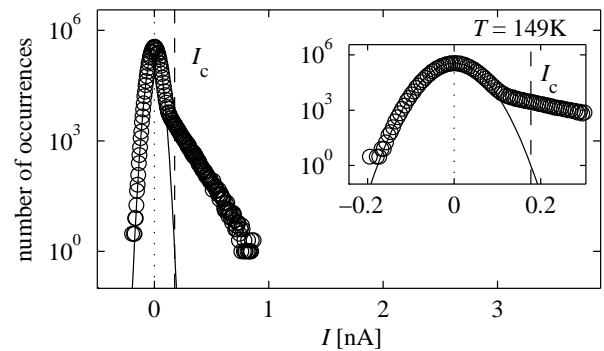}
\caption{Histogram of measured current values for $T=192$~K after
  shifting the maximum to $I=0$. The solid line marks a Gaussian fit
  with zero mean and standard deviation determined from the histogram
  for negative currents. The inset shows an enlargement to demonstrate
  the determination of the threshold value $I_c$ used for separating
  the diffusion-induced signal from the noise (see text).}
\label{fig:noiselevel}
\end{figure}

\subsection*{Application to CuPc on Ag(100)}

A relaxed structure of the CuPc molecule in vacuum after energy
minimization, as obtained from a DFT calculation, is displayed in
Fig.~\ref{fig:cupc}, together with. A circle of radius $R=7.6$~\AA\ is
assigned to it.  The relaxed structure can be expected to be
essentially unmodified when the CuPc molecule is adsorbed on the
Ag(100) surface. STM studies combined with DFT calculations showed
that transition metal phtalocyanines generally lie flat on the Ag(100)
surface.\cite{Mugarza/etal:2012} This is in agreement with STM images
of CuPc islands on Ag(100).\cite{Bach:2009}

After deposition of CuPc molecules on Ag(100) up to a coverage of
10-15\%, islands form and a quasi-stationary state is reached, where
the rates of detachment from and attachment to islands balance each
other. In this quasi-stationary state the diffusing CuPc molecules
have an effective small number density $c$.  Then the tunneling
current was measured for a stationary STM tip and constant bias
voltage. The height of the tip was preset indirectly by the choice of
setpoint current $I_{\textrm{setpoint}}$ and bias voltage
$U_{\textrm{bias}}$. Further details of the experiment are given in
Ref.~\onlinecite{Ikonomov/etal:2010}.

\begin{figure}[t!]
\includegraphics[width=8cm]{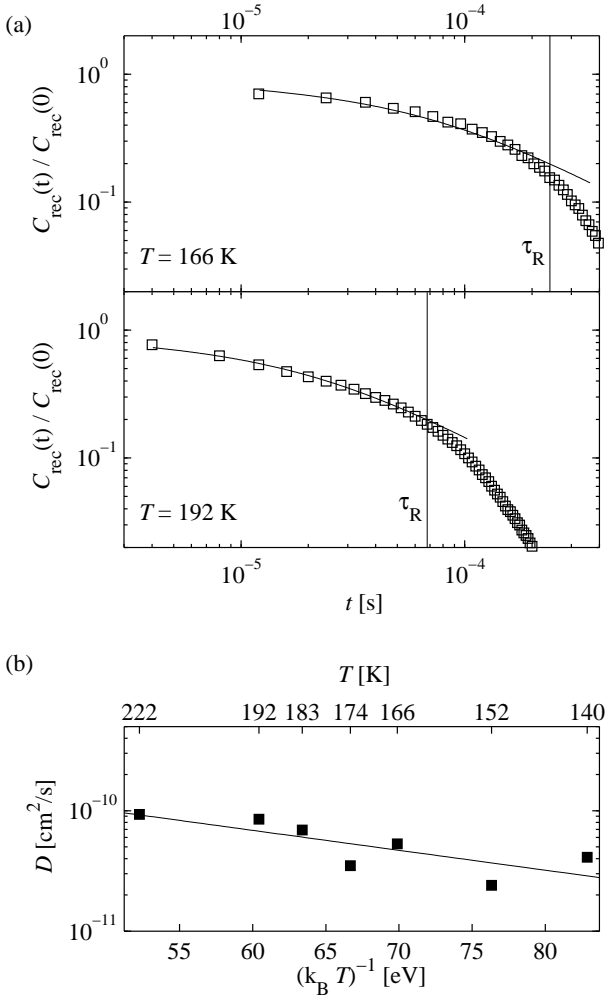}
\caption{(a) Normalized ACFs $C_\textrm{rec}(t)/C_\textrm{rec}(0)$ as
  a function of time for two temperatures (symbols) and best fits with
  Eq.~(\ref{eq:krec}) for $t<\tau_R$ (solid lines). (b) Arrhenius plot
  of the resulting diffusion coefficients (symbols). A least square
  fit (solid lines) according to the Arrhenius law yield $E_a=30$~meV
  and $D_0=10^{-9.4}$~cm$^2$/s.}
\label{fig:acf-fit}
\end{figure}

The diffusing molecules cause fluctuations in the tunneling current
$I(t)$ as shown in Fig.~\ref{fig:currents} (a) for a time interval of
2~ms at $T=192$~K. A histogram of the tunnelling current, determined
for the whole time series of length 40~s, is depicted in
Fig.~\ref{fig:noiselevel}. In this figure, the maximum of the
histogram was shifted to $I=0$.\cite{comm:ishift} Negative $I$ values
are associated with noise and their distribution can be fitted to a
half-sided Gaussian. Extending this Gaussian distribution to positive
$I$ values yields the solid line in Fig.~\ref{fig:noiselevel}. For
small positive $I$ values this curve fits very well the data, implying
that these values can also be attributed to noise. The diffusing CuPc
molecules cause deviations from the Gaussian distribution at larger
$I$ values. To separate the diffusion-induced fluctuations from the
noise, we define a threshold value $I_c>0$, where the Gaussian
histogram of the number of occurrences drops below one, cf.\
Fig.~\ref{fig:noiselevel}.  This threshold is marked by the dashed
line in Fig.~\ref{fig:currents}(a) and the corresponding
$I_\textrm{rec}(t)$ is shown in Fig.~\ref{fig:currents}(b). Using this
procedure we determined the ACF of $I_\textrm{rec}(t)$ for seven
different temperatures investigated in the
experiments.\cite{Ikonomov/etal:2010}

Normalized ACFs $C_\textrm{rec}(t)/C_\textrm{rec}(0)$ for two
temperatures are shown in Fig.~\ref{fig:acf-fit}(a) (symbols) and fits
to them with Eq.~(\ref{eq:krec}) and under the constraint $t<\tau_R$
are marked as solid lines. The resulting diffusion coefficients are
displayed in the Arrhenius plot in Fig.~\ref{fig:acf-fit}(b)
(symbols), where a least square fit of
$D=D_0\exp(-E_a/k_{\textrm{B}}T)$ to these data (solid line) yields an
activation energy $E_a=30$~meV and a pre-exponential factor
$D_0=10^{-9.4}$~cm$^2$/s.

\section{Residence time distribution (RTD)}\label{sec:rtd}

The residence times $\tau$ are the time intervals between entrance and
exit of a molecule into the detection region $\mathcal{M}$ of the
probe, see Fig.~\ref{fig:currents}. By sampling these $\tau$ values,
the RTD $\Psi(\tau)$ is obtained.

\begin{figure}[t!]
\includegraphics[width=8cm]{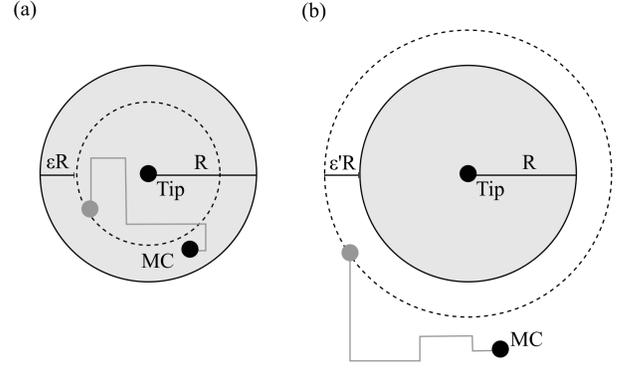}
\caption{Illustration of the geometry used for the calculation of (a)
  the RTD and (b) the ITD. The dashed lines indicate the uniformly
  distributed initial probabilities on the rings displaced by
  $\epsilon R$ (a) and $(\epsilon'R)$ (b) from the absorbing boundary
  (solid lines).  MC stands for the diffusing molecule center.}
\label{fig:sketch}
\end{figure}

For a theoretical description of the RTD we need to tackle the problem
of the diffusion of a molecule center in a circle with radius $R$ and
absorbing boundaries, see Fig.~\ref{fig:sketch}a. Since the molecule
center cannot be placed at the absorbing boundary, its initial center
position is considered to be at a distance $\epsilon R$ from the
boundary.  This means that the initial probability density of the
center position is a delta-function concentrated on the ring with
radius $(1-\epsilon)R$. Physically the length $\epsilon R$ may be
associated with an elementary step size of the molecule on the
substrate.

Due to the symmetric initial condition, the probability density for
the molecule center position $\mathbf{r}$ at time $t$ depends on
$r=|\mathbf{r}|$ only and is given by (see Appendix)
\begin{equation}
p(\mathbf{r},t)=\sum_{n=1}^\infty \frac{J_0(\alpha_n r/R)}{\pi R^2}
\frac{J_0(\alpha_n (1-\epsilon))}{J_1^2(\alpha_n)}\,e^{-\alpha_n^2 t/\tau_R}\,,
\label{eq:pxyt}
\end{equation}
where $J_k(.)$ are the Bessel functions of order $k$, and the
$\alpha_n$ are the (positive) zeros of $J_0$, $J_0(\alpha_n)=0$
with $0<\alpha_1<\alpha_2<\ldots$

With $p(\mathbf{r},t)$ from Eq.~(\ref{eq:pxyt}), the calculation of
the RTD amounts to calculating the first passage time distribution
with respect to the absorbing boundary. This can be carried out by
applying standard techniques from random walk theory, as described,
for example, in Ref.~\onlinecite{Redner:2001}. The probability that
the molecule center has not left the circle until time $t$ is
$\phi(t)=2\pi\int_0^R\textrm{d}r\,rp(\mathbf{r},t)$. The probability
that the center leaves the circle in the time interval
$[\tau,\tau+\Delta\tau]$ is $\phi(\tau)-\phi(\tau+\Delta\tau)$,
implying that the probability density $\psi(\tau)$ for the residence
times $\tau$ is $\psi(\tau)=-2\pi\partial_\tau\int_0^R\textrm{d}r\,r
p(\mathbf{r},\tau)$. This yields
\begin{equation}
\psi(\tau)=
\frac{2}{\tau_R}\sum_{n=1}^\infty \alpha_n
\frac{J_0(\alpha_n (1-\epsilon))}{J_1(\alpha_n)}\,
\textrm{e}^{-\alpha_n^2 \tau/\tau_R}\,.
\label{eq:psi}
\end{equation}
We note that the characteristic time $(\epsilon R)^2/D$ sets a lower
limit above which Eq.~(\ref{eq:pxyt}) becomes applicable, because one 
would have to refine the continuum treatment  for smaller $\tau$. For
the experiments analyzed here, this is not of relevance because such
small $\tau$ are not resolved.

For $\tau\gtrsim(\epsilon R)^2/D$ the functional form of the solution
changes with respect to the characteristic time $\tau_R$
the molecule needs to explore the circle area. For $\tau\ll\tau_R$,
we find
\begin{equation}
\psi(\tau)\propto \frac{1}{\tau_R}
\left(\frac{\tau}{\tau_R}\right)^{-3/2}\,.
\label{eq:psi1}
\end{equation}
This power law can be understood by considering the (negative) time
derivative of the probability that the molecule is next to the
absorbing boundary at time $t$ [i.e.\ the efflux rate, which equals
$\psi(\tau)$]. For $\tau\ll\tau_R$ this probability is given by the
ratio of the explored boundary section ($\propto\sqrt{\tau}$) to the
explored area inside the circle ($\propto \tau$), yielding
$\Psi(\tau)\sim-\partial_{\tau} \sqrt{\tau}/\tau\sim\tau^{-3/2}$.

For $\tau\gg\tau_R$, we find
\begin{equation}
  \psi(\tau)\sim \frac{2}{\tau_R}
\frac{\alpha_1J_0(\alpha_1(1-\epsilon))}{J_1(\alpha_1)}\,
\textrm{e}^{-\alpha_1^2\tau/\tau_R}\,.
\label{eq:psi2}
\end{equation}
This result can be interpreted by noting that for $\tau\gg\tau_R$,
where the occupation probability of the molecule is spread over the
circle, the efflux rate is essentially constant and given by the
inverse of the time $\tau_R$ for a molecule to reach the boundary.  A
Poisson process with this constant rate yields
$\Psi(\tau)\sim\tau_R^{-1}\exp(-const.~\tau/\tau_R)$

\subsection*{Application to measurements}

\begin{figure}[t!]
\includegraphics[width=8cm]{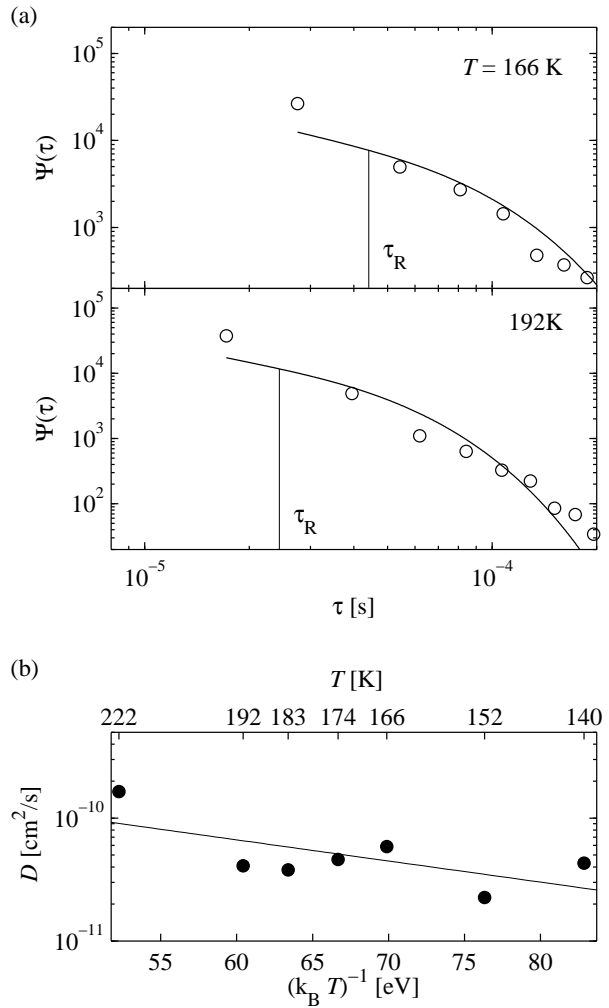}
\caption{(a) Residence time distribution for CuPc/Ag(100) at two
  temperatures (symbols). The solid line marks the result predicted by
  Eq.~(\ref{eq:pxyt}) after determining the parameters $D$ and
  $\epsilon$ from the exponential tail region $\tau\gg\tau_R$ (cf.\
  Eq.~(\ref{eq:psi2}) and discussion in text). (b) Arrhenius plot of
  the extracted diffusion coefficients. A least square fit (solid
  line) with the Arrhenius law yields an activation energy
  $E_a=30$~meV and a pre-exponential factor $D_0=10^{-9.4}$~cm$^2$/s.}
\label{fig:RTD}
\end{figure}

For the diffusion of CuPc on Ag(100) the $\tau$ values were extracted
from the rectangular current signal as described in
Fig.~\ref{fig:currents}. Distributions $\Psi(\tau)$ of these $\tau$
values are shown in Fig.~\ref{fig:RTD}(a) for two representative
temperatures. By fitting the exponential tail with
Eq.~(\ref{eq:psi2}), we obtain the diffusion coefficient $D$. In a
self-consistency check we have assured that the tail regime used for
the fitting fulfills the requirement $\tau\gg\tau_R=R^2/D$. In
addition the value $\epsilon$ can be calculated from the prefactor of
the fit. We find $\epsilon\simeq 0.3$, which means that $\epsilon R$
is about the lattice constant $a=2.89$~\AA\ of the Ag(100)
substrate. Inserting $D$ and $\epsilon$, Eq.~(\ref{eq:psi}) yields the
full distribution $\psi(\tau)$, which is marked by the solid lines in
Fig.~\ref{fig:RTD}(a). The very good agreement of the theoretical
prediction with the measured data demonstrates the reliability of the
approach.

The diffusion coefficient $D$ is shown in Fig.~\ref{fig:RTD}(b) for
the two temperatures in (a) together with the five further
investigated temperatures in an Arrhenius plot. From the slope of the
fitted line, we find an activation energy $E_a=30$~meV and a
pre-exponential factor $D_0=10^{-9.4}$~cm$^2$/s. These values agree
with the ones of the ACF analysis. 

We note, however, that these values deviate from those reported in
Ref.~\onlinecite{Ikonomov/etal:2010} ($E_a=81$~meV and
$D_0=10^{-8.4}$~cm$^2$/s).  The difference is due to the fact that
error bars were taken into account in the analysis performed in
Ref.~\onlinecite{Ikonomov/etal:2010}. One point in the Arrhenius plot,
which was determined from the data set measured for $T=222$~K, had a
particularly small error and was largely influencing the slope of the
fit line because of its exposed position with respect to the other
points. When excluding this particular point from the fitting, values
$E_a=38$~meV and $D_0=10^{-9.7}$~cm$^2$/s are obtained, in fair
agreement with the present analysis. We have refrained from including
error bars here because the fits of the RTD for different temperatures
yielded comparable errors and small differences between them seem to
be insignificant with respect to other possible sources of errors, as,
for example, minor temperature fluctuations or the influence of
spatial inhomogeneities in diffusion profiles, that are associated
with the fact that islands act as sinks for the molecule diffusion.

\section{Inter-peak time distribution (ITD)}\label{sec:itd}

The inter-peak times $\tau'$ are the time intervals between
the end of a peak and the beginning of the next peak in the
rectangular signal, see Fig.~\ref{fig:currents}.  The statistics of
them is, for small $\tau'$, dominated by entrance and exit of the same
molecule into the detection region $\mathcal{M}$.  For calculating the
contribution of these return processes to the ITD we analyze the
diffusion of a molecule center with initial distance $\epsilon'R$ from
a circular absorbing boundary with radius $R$, see Fig.~\ref{fig:sketch}. 
For the probability density of the molecule center to be
at position $\mathbf{r}$ at time $t$ one obtains (see Appendix)
\begin{align}
\label{eq:pitd}
p(\mathbf{r},t)&=
\int_{0}^{\infty}\frac{\textrm{d}\alpha}{2\pi R^2}\, 
\alpha\,W_0(\alpha r/R,\alpha)\\
&\hspace{5em}
{}\times\frac{W_0(\alpha(1+\epsilon'),\alpha)}
{J_0^2(\alpha)+Y_0^2(\alpha)}\,\textrm{e}^{-\alpha^2t/\tau_R}\,,
\nonumber
\end{align}
where $W_0(x,y)=J_0(x)Y_0(y)-J_0(y)Y_0(x)$ and $Y_0$ is the
zeroth order Bessel function of second kind.

The ITD can be derived analogously to the treatment of the RTD by taking
the time derivative of the integral of $p(\mathbf{r},t)$ over the outer
area with respect to the circle. In the present case it is
more convenient to take the flow through the absorbing boundary,
$\psi(\tau')=\oint\textrm{d}\mathbf{s}
\cdot [-D\mathbf{\nabla} p(\mathbf{r},t)]_{r=R}$, which, when
making use of the Wronskian $[J_0'(x)Y_0(x)-J_0(x)Y_0'(x)]=2/\pi x$,
yields
\begin{align}
\label{eq:psiitd}
\psi(\tau')=&
\frac{2}{\pi \tau_R}
\int_{0}^{\infty}\textrm{d}\alpha\,\alpha\, 
\frac{W_0(\alpha(1+\epsilon'),\alpha)}
{J_0^2(\alpha)+Y_0^2(\alpha)}\,\textrm{e}^{-\alpha^2\tau'/\tau_R}\,.
\end{align}
For $\tau'\ll\tau_R$, the asymptotic behavior for $\tau'\to0$ is
\begin{equation}
\psi(\tau')\sim\frac{\epsilon'}{2\tau_R\sqrt{\pi (1+\epsilon')}}
\exp\left(-\frac{\epsilon'^2\tau_R}{4\tau'}\right)
\left(\frac{\tau'}{\tau_R}\right)^{-3/2}\,.
\label{eq:psiitd1}
\end{equation}
Accordingly, $\psi(\tau')$ rapidly rises for small $\tau'$ and, after
going through a maximum $\psi_{\textrm{max}}$ at
$\tau'_{\textrm{max}}=(\epsilon'R)^2/6D$, approaches a power law with
exponent (-3/2). This power law has an analogous physical origin as
the power law in the RTD, see the discussion in Sec.~\ref{sec:rtd}
after Eq.~(\ref{eq:psi1}). Here $\tau_R$ is the typical time where the
molecule center in Fig.~\ref{fig:sketch} realizes the finite extent of
the detection area, or, in other words, where the molecules realizes
its size.

For $\tau'\gg\tau_R$, Eq.~(\ref{eq:psiitd}) can be approximated by
\begin{equation}
\psi(\tau')\simeq
\frac{2\ln(1+\epsilon')}{\tau'\ln^2(\tau'/\tau_R)}.
\label{eq:psiitd2}
\end{equation}
The asymptotics $\sim(\tau\ln^2\tau)^{-1}$ follows from the fact that
for large $\tau'$ the detection area becomes very small with respect
to the area explored by the molecule.  Accordingly, $\Psi(\tau')$
scales as the probability of first return to the origin of a
two-dimensional random walk.

The large $\tau$ behavior predicted by Eq.~(\ref{eq:psiitd2}) is,
however, of limited use, because another molecule can enter the
detection area before the molecule, which has left this area at last,
returns to it. The memory to a molecule that leaves the detection area
is lost on time scales of order $l^2/D\sim 1/cD$.  On these time
scales different molecules can be regarded as entering the detection
area with a constant rate. This rate should scale with the inverse
mean time $D/l^2$ for a molecule outside the detection area to enter
it. Hence in the limit of large $\tau'$, an exponential distribution
is expected,
\begin{equation}
  \psi(\tau')\sim cD\exp\left(-\kappa \pi Dc\tau'\right)\,,
\label{eq:psiitd3}
\end{equation}
where $\kappa$ is a constant of order unity.

\begin{figure}[t!]
\includegraphics[width=8cm]{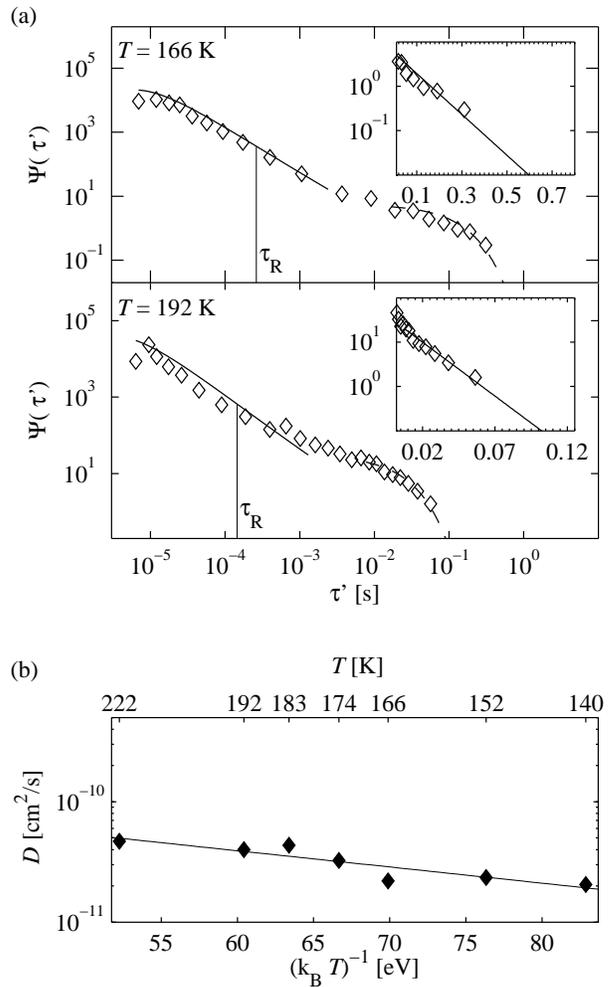}
\caption{(a) ITD for CuPc/Ag(100) for two temperatures.  The inset
  shows the exponential decay at large times, where the straight solid
  lines (dashed lines in the main log-log plots) are fits with
  Eq.~(\ref{eq:psiitd3}), yielding $D$ (see text). Using these $D$
  values, the solid lines in the main plots mark the short time
  behavior after fitting with Eq.~(\ref{eq:psiitd1}). (b) Arrhenius
  plot of the extracted diffusion coefficients for all seven
  investigated temperatures.  A least square fit (solid line) with the
  Arrhenius law yields an activation energy $E_a=31$~meV and a
  pre-exponential factor $D_0=10^{-9.6}$~cm$^2$/s.}
\label{fig:ITD}
\end{figure}

\begin{table*}
\caption{\label{tab:summary} 
Strengths and weaknesses of the three methods for determining $D$.}
\begin{ruledtabular}
  \begin{tabular}{p{.15\textwidth}p{0.05em}p{.21\textwidth}p{0.05em}
      p{.21\textwidth}p{0.05em}p{.21\textwidth}}
    Characteristics & ACF & & RTD & & ITD & \\
    \hline \hline
    Signal processing & $\oplus$ & Convenient by FFT & $\ominus$ & 
    Peak widths affected by $I_c$ value & $\oplus$ & 
    Negligible influence of $I_c$ on interpeak distances \\
    \hline 
    Tip influence & $\ominus$ & Possible & $\ominus$ & 
    Possible & $\oplus$ & Less likely \\
    \hline 
    Assumptions & $\ominus$ & 
    Non-interacting particles & $\oplus$ & None  & $\oplus$ & 
    None in short-time regime \\
    &   &   &   &    & $\ominus$ & Non-interacting particles
 in long-time regime \\
\end{tabular}
\end{ruledtabular}
\end{table*}

\subsection*{Application to measurements}
The inter-peak intervals were sampled from the current signal as
explained in Fig.~\ref{fig:currents} and the ITDs for two temperatures
are shown in Fig.~\ref{fig:ITD}(a). Note that for very small times the
data indicate a plateau, which is not contained in
Eq.~(\ref{eq:psiitd1}). A refinement of the continuum description
would be necessary to capture this behavior, analogous to the very
small times in the RTD.  For times $\tau\gtrsim\tau_R$ the power law
predicted by Eq.~(\ref{eq:psiitd1}) is reflected by the straight line
behavior in the double-logarithmic plot, before eventually the
exponential decay according to Eq.~(\ref{eq:psiitd3}) takes over.

To obtain $D$ from the ITD, we first concentrate on the long-time
behavior. Setting $\kappa=1$, the rate $cD$ of the exponential decay
is provided by the slopes of the lines shown in the insets of
Fig.~\ref{fig:ITD}(a).  The concentration is determined from the
probability $c\pi R^2=\sum_i\tau_i/T$ for a molecule to be in the
detection area, where $T=\sum_i(\tau_i+\tau'_i)$ is the total
measurement time. For the seven data sets taken at different
temperatures, we find $c$ values in the range
$0.6-5\times10^{-6}\,\mbox{\AA}^2$, corresponding to coverages
$\Theta=0.01-0.08\%$ of the diffusing CuPc molecules. The resulting
$D$ values are shown in the Arrhenius plot in Fig.~\ref{fig:ITD}(b)
and yield an activation energy of $E_a=31$~meV and a pre-exponential
factor of $D_0=10^{-9.6}$~cm$^2$/s. Note that the assumed value
$\kappa=1$ affects only the prefactor $D_0$ but not the activation
energy $E_a$.

With $D$ obtained from the long-time behavior, we can fit the
remaining part of the ITD with Eq.~(\ref{eq:psiitd1}). Corresponding
curves are shown as solid lines in the double-logarithmic plots of
Fig.~\ref{fig:ITD}(a). They give a good agreement with the experimental
data. The fits yield $\epsilon'\simeq0.5$, which is consistent
with $\epsilon\simeq0.3$ obtained in the analysis of the RTD, see
Sec.~\ref{sec:rtd}. Accordingly, we find again that $\epsilon'R$ is
of the order of the lattice constant $a$, as one should expect.

In principle, the part of the ITD dominated by the single molecule
diffusion can also be used to determine $D$. For this we have to notice
that the necessary refinements of the continuum theory to describe the
behavior of the ITD left to the maximum 
\begin{equation}
\psi_{\textrm{max}}=
\frac{3\sqrt{6}D}{\sqrt{\pi(1+\epsilon')\textrm{e}^{3/2}(\epsilon' R)^2}}
\approx \frac{D}{a^2}.
\label{eq:psiitd0}
\end{equation}
are not expected to yield larger values of the ITD. In fact, when
considering jump dynamics of the molecules with a rate $D/a^2$ for
short times, the ITD should behave as $\sim
(D/a^2)\exp(-const.~D\tau'/a^2)$, i.e.\ the largest value of the ITD
should be of order $D/a^2$. Due to matching with the continuum
treatment, we can identify $\psi_{\textrm{max}}$ with the maxima seen
in Fig~\ref{fig:ITD}(a). This then is a convenient way to determine
$D/(\epsilon' R)^2$, and knowing this value, to extract $D$ by fitting
the part right to the maximum predicted by
Eq.~(\ref{eq:psiitd1}). Application of this alternative method indeed
yields values for $D$ and $\epsilon'$ in good agreement with those
discussed above.

\section{Comparison of the methods}
\label{sec:comb}

For applications it is important to clarify how the three methods are
best combined to obtain most accurate results for $D$.  To this end we
need to evaluate the strengths and weaknesses of each method.  Let us
first note that all methods work with a single input parameter, which
is the molecule radius $R$. All other variables arise from the
analyses described in Secs.~\ref{sec:acf}-\ref{sec:itd}. Because
$\tau_R=R^2/D$ is actually determined, uncertainties with respect to
$R$ slightly affect the diffusion coefficient.

Table~\ref{tab:summary} summarizes the advantages and disadvantages of
the three methods. The ACF can be readily calculated by a fast Fourier
transformation without caring about peak identification in the
signal. A disadvantage is that only the short-time regime $t\lesssim
\tau_R$ is governed by single particle diffusion, while an accurate
theoretical description of the crossover to the long-time regime,
governed by collective particle diffusion, requires a careful
consideration of the mutual exclusion of the molecules (and
possibly other interaction effects). The decay of the correlation
function within this regime can be small, which then affects the
accuracy of the $D$ values resulting from the fitting.  Another
drawback is that the determination of the ACF includes time intervals
where molecules are under the tip and possible interactions with the
tip can thus have an influence on the diffusion properties.

Both the RTD and ITD method require some more effort in preparation of
the data, which is connected to the determination of the threshold current
$I_c$ and identification of the peaks. Once $I_c$
is set, both the peak widths and interpeak distances can be extracted
simultaneously. Note that any method of determining $I_c$ is
associated with some uncertainty. For the ITD this is no problem in
practice, because the interpeak intervals are comparatively large. 
We have checked that, when taking the interpeak
distances as time intervals between peak maxima, the results do not
change significantly. For the RTD the uncertainty of $I_c$ is a more severe
problem. Because molecules diffuse slowly into the detection area,
the peaks in the original tunneling current signal have rather flat
flanks. As a result the peak widths change more sensitively with $I_c$ 
than the interpeak intervals.

An advantage of the RTD is that $D$ can be determined solely by
analyzing the exponential tail for large residence times. One should
note, however, that it may be difficult to obtain a good statistics in
the tail regime, when the molecules are highly mobile or small. In
this case the peaks are narrow and it could be difficult to resolve
them accurately.  In the RTD method the interaction with the tip can
influence the residence times and in this case one would not determine
the free diffusion of the molecules on the substrate. By
systematically changing the bias voltage, a possible influence can be
reduced to a minimum.\cite{Ikonomov/etal:2010} A strength of the RTD
is that it is related to a single-particle problem.

The ITD method has the advantage that tip-molecule interactions can be
expected to have, if at all, a marginal influence on the interpeak
times (irrespective of tip-substrate distances or bias voltages). For
small interpeak times, the ITD is essentially also related to a
single-particle problem.  For large interpeak times, an approximate
value of $D$ can be obtained based on an estimate for processes, where
one molecule in the detection area under the tip is followed by
another molecule. The values obtained in this way turn out to be close
to those resulting from the other methods. Compared to the residence
times, the interpeak times are quite large and are thus less prone to
the experimental time resolution and the threshold value $I_c$.

\section{Conclusions}\label{sec:conclusions}

\begin{figure}[t!]
\includegraphics[width=8cm]{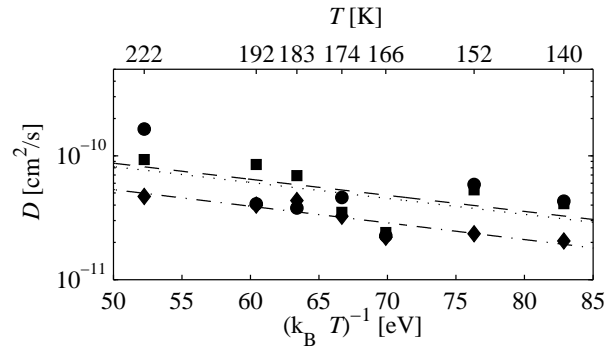}
\caption{Comparison of the diffusion coefficients obtained from the
  three methods. Squares, dashed line: ACF method ($E_a = 30$~meV,
  $D_0 = 10^{-9.4}$~cm$^2$/s); Circles, dotted line: RTD method ($E_a
  = 30$~meV, $D_0 = 10^{-9.4}$~cm$^2$/s); Diamonds, dashed-dotted
  line: ITD method ($E_a = 31$~meV, $D_0 = 10^{-9.6}$~cm$^2$/s).}
\label{fig:arrhenius}
\end{figure}

In order to exploit the strength of each method one can combine the
different methods, if the corresponding data in the relevant time
regime have sufficient statistics. For a first classification of the
measurement it is helpful to choose the ACF method, because it yields
fast results without analyzing the time series in detail. In a
subsequent step the peak widths and interpeak intervals should be
extracted with the procedure described in Sec.~\ref{sec:acf}. We then
recommend to use the RTD method for determining $D$ whenever the
exponential tail is sufficiently pronounced and the tip influence on
the diffusion can be neglected.  Otherwise the ITD method should be
preferred. In any case, one should always apply both the RTD and the
ITD method to perform consistency checks and to obtain most reliable
results. Our analysis for CuPc on Ag(100) shows that, despite
diffusion coefficients extracted from the different methods may differ
for one or another sample, the activation energies determined from the
Arrhenius plots should have comparable values, see
Fig.~\ref{fig:arrhenius}.

An interesting question is whether the methods discussed here can be
taken over to other fields. Obviously, the ACF method is well known in
fluorescence correlation spectroscopy
(FCS),\cite{Petrov/Schwille:2008} where intensity fluctuations reflect
concentration fluctuations, typically in some finite detection volume.
Similarly, analysis of density fluctuations has been applied as one
variant to determine diffusion coefficients in Field Emission
Microscopy.\cite{Gomer:1990} In the situation discussed in
Sec.~\ref{sec:acf} only one particle is in the detection area and
accordingly information on the tagged particle diffusion (tracer
diffusion) is obtained.  In the hypothetical limit of very small
particle concentrations in FCS, where the mean interparticle distance
becomes larger than the linear size of the detection volume, one
should essentially recover the behavior discussed in
Sec.~\ref{sec:acf} (with $R$ then playing the role of the detection
length).  In common applications of this method one is, however, not
able to extract the tagged particle information and this makes a
difference to the ACF method discussed in Sec.~\ref{sec:acf}. This
also prevents the use of the RTD and ITD in the analysis of typical
fluorescence signals. These methods can yet become useful in light of
the ongoing development of sophisticated techniques to measure
single-molecule properties.

In summary, we provided formulas for different means of analyzing
fluctuations in the signal of a locally fixed probe on a surface,
where the fluctuations are caused by diffusing particles. By
exploiting the fact that the molecule extension enters as a scale bar
into the equations, a quantitative determination of the diffusion
constant becomes possible.  So far, we have considered circular shapes
for planar molecules. The situation will evidently be more involved
when rectangular or even more complicated shapes are considered. In
addition rotational diffusion and anisotropies induced by the
substrate then can modify the results. As mentioned above, the
interaction of the STM tip with the diffusing molecule was neglected
so far. The interaction is influenced by the experimental setup (tip
shape, tip-substrate distance etc.) and it is specific for the tip and
molecule material. One objective could be to minimize the tip
influence experimentally. On the other hand, it could be interesting
to study how the interaction potential affects the diffusion of the
molecule. This potential has so far not been included in the analysis,
but it should be possible and bears an interesting route to obtain
information on the tip-molecule interaction.

\begin{acknowledgments}
  J.I. and M.S. gratefully acknowledge fincancial support by the SFB
  624 of the Deutsche Forschungsgemeinschaft.
\end{acknowledgments}

\appendix*
\section{Diffusion propagators for RTD and ITD}
\label{sec:appendix}
The results given in Eqs.~(\ref{eq:pxyt}) and (\ref{eq:pitd}) for
$p(\mathbf{r},t)$ have been derived earlier in the literature, in
particular Eq.~(\ref{eq:pitd}) by using the Heaviside
method.\cite{Goldstein:1932} We give a straightforward derivation
here, using separation of variables and eigenfunction expansions.

Let us first consider the equation for diffusion of a particle in a
circular stripe $a<r<c$ ($a$ here not equal to the lattice
constant used in the main text) with absorbing boundaries and initial
distribution
\begin{equation}
  p_0(\mathbf{r}) =  \frac{1}{2\pi r} \delta(r-b)\,,\ a<b<c\,.
  \label{eq:a.1}
\end{equation}
Due to the radial symmetry $p(\mathbf{r},t) = \rho(r,t)/2\pi$, where
$\rho(r,t)$ satisfies the radial diffusion equation
\begin{equation}
  \frac{\partial\rho(r,t)}{\partial t} = 
D\left(\frac{\partial^2}{\partial r^2}+\frac{1}{r}
\frac{\partial}{\partial r}\right)\rho(r,t)
  \label{eq:a.2}
\end{equation}
with $\rho(a,t)=\rho(c,t)=0$ and $\rho(r,0)=\delta(r-b)/r$.

Making the product ansatz $\rho(r,t)=\rho(r,0)g(t)$,
Eq.~(\ref{eq:a.2}) separates in the variables $r$ and $t$. One obtains
$g(t)= \textrm{e}^{-\lambda^2 Dt}$, where the allowed values for
$\lambda^2$ are the eigenvalues of the radial Laplace operator in the
circular stripe,
\begin{equation}
  \left(\frac{\partial^2}{\partial r^2}+
\frac{1}{r}\frac{\partial}{\partial r}\right) \Psi_n(r) 
= -\lambda^2_n \Psi_n(r)\,,\ \Psi_n(a)=\Psi_n(c)=0\,.
  \label{eq:a.3}
\end{equation}
Because the Laplacian is negative definite, $\lambda_n^2>0$.  The
eigenfunctions $\Psi_n$ are given by linear combinations of the zeroth
order Bessel functions $J_0(.)$ and $Y_0(.)$ of first and second kind,
$\Psi_n(r) = A_n J_0(\lambda_n r) + B_n Y_0(\lambda_n r)$ (where
restriction to $\lambda_n>0$ gives linear independent
eigenfunctions). The Dirichlet boundary conditions yield $A_n
J_0(\lambda_n a) = -B_n Y_0(\lambda_n a)$ [or $A_n J_0(\lambda_n c) =
-B_n Y_0(\lambda_n c)$], and
\begin{equation}
  J_0(\lambda_n a)Y_0(\lambda_n c) - J_0(\lambda_n c)Y_0(\lambda_n a) = 0
  \label{eq:a.4}
\end{equation}
as determining equation for the $\lambda_n$, $n=1,2,\ldots$
($0<\lambda_1<\lambda_2<\ldots$). Introducing the cross product
\begin{equation}
  W(x,y) = J_0(x)Y_0(y)-J_0(y)Y_0(x)\,,
  \label{eq:a.5}
\end{equation}
the solution becomes
\begin{equation}
  \rho(r,t) = 
\sum_{n=1}^\infty C_n W_0(\lambda_n r, \lambda_n a) \textrm{e}^{-\lambda_n^2 Dt}\,,
  \label{eq:a.6}
\end{equation}
where $W_0(\lambda_n a, \lambda_n c) = 0$. Utilizing the orthogonality
of the eigenfunctions,
\begin{eqnarray}
  \int_a^c \textrm{d}r\ r\ 
W_0(\lambda_m r,\lambda_m a)W_0(\lambda_n r, \lambda_n a)\nonumber\\
    =\delta_{mn} \int_a^c \textrm{d}r\ r\ W_0^2(\lambda_n r,\lambda_n a)\,,
  \label{eq:a.7}
\end{eqnarray}
the expansion coefficients $C_n$ follow from the initial condition
Eq.~(\ref{eq:a.1}):
\begin{eqnarray}
  C_n &=& 
\frac{\int_a^c \textrm{d}r\ r\ W_0(\lambda_n r,\lambda_n,a)\rho(r,0)}
{\int_a^c \textrm{d}r\ r\ W_0^2(\lambda_n r,\lambda_n a)}\nonumber\\
  &=& \frac{W_0(\lambda_n b,\lambda_n a)}
{\int_a^c \textrm{d}r\ r\ W_0^2(\lambda_n r,\lambda_n a)}\,.
  \label{eq:a.8}
\end{eqnarray}
The result for $p(\mathbf{r},t)$ thus is
\begin{equation}
  p(\mathbf{r},t) = \sum_{n=1}^\infty 
\frac{W_0(\lambda_n b,\lambda_n a)W_0(\lambda_n r,\lambda_n a)}
{2\pi \int_a^c \textrm{d}r\ r\ W_0^2(\lambda_n r, \lambda_n a)} 
\textrm{e}^{-\lambda_n^2 Dt}\,.
  \label{eq:a.9}
\end{equation}
The diffusion propagator in Eq.~(\ref{eq:pitd}) relevant for the ITD
corresponds to the limit $c\to\infty$, where the spectrum of
eigenvalues determined by Eq.~(\ref{eq:a.4}) becomes
continuous. Analogous to the change of a Fourier series to a Fourier
integral, we are led to consider the Weber transform\cite{Davies:1978}
of a function $f=f(r)$
\begin{equation}
  F(\lambda) = \frac{1}{a^2}\int_a^\infty \textrm{d}r \,
r\, W_0(\lambda r, \lambda a)f(r)\,,
  \label{eq:a.10}
\end{equation}
with back-transformation
\begin{equation}
  f(r) = a^2 \int_0^\infty \textrm{d}\lambda\,\lambda\,
\frac{W_0(\lambda r,\lambda a)}{J_0^2(\lambda a)+Y_0^2(\lambda a)}F(\lambda)\,.
  \label{eq:a.11}
\end{equation}
Accordingly, Eq.~(\ref{eq:a.6}) becomes
\begin{equation}
  \rho(r,t) = \int_0^\infty \textrm{d}\lambda\, 
  C(\lambda) W_0(\lambda r,\lambda a)\,\textrm{e}^{-\lambda^2 Dt}\,,
  \label{eq:a.12}
\end{equation}
where
\begin{eqnarray}
  C(\lambda) &=& \frac{\lambda}{J_0^2(\lambda a)+Y_0^2(\lambda a)} 
\int_a^\infty \textrm{d}r 
r W_0(\lambda r,\lambda a)\rho(r,0)\nonumber\\
  &=& 
\frac{\lambda W_0(\lambda b, \lambda a)}{J_0^2(\lambda a)+Y_0^2(\lambda a)}\,.
  \label{eq:a.13}
\end{eqnarray}
This yields
\begin{equation}
  p(\mathbf{r},t) = \int_0^\infty \frac{\textrm{d}\lambda}{2\pi}\,\lambda\,
\frac{W_0(\lambda r,\lambda a)W_0(\lambda b,\lambda a)}
{J_0^2(\lambda a)+Y_0^2(\lambda a)}\,\textrm{e}^{-\lambda^2 Dt}\,.
  \label{eq:a.14}
\end{equation}
Equation (\ref{eq:pitd}) follows by setting $a=R, b=(1+\epsilon)R$,
and $\lambda=\alpha/R$.

For the diffusion propagator in Eq.~(\ref{eq:pxyt}) relevant for the
RTD only one boundary condition $\rho(c,t)=0$ has to be taken into
account. In this case the Bessel functions of second kind cease to
apply, because their logarithmic singularity at the origin eliminates
them from the space of functions, where the radial Laplace operator is
Hermitean. Notice that the logarithmic singularity would be no problem
with respect of the integrability of $p(\mathbf{r},t)$. The
eigenfunctions thus are given by $\Psi_n(r) = A_n J_0(\lambda_n r)$,
where the $\lambda_n$ are determined by $J_0(\lambda_n c) =
0$. Equation (\ref{eq:a.6}) becomes
\begin{equation}
  \rho(r,t) = \sum_{n=1}^\infty A_n J_0(\lambda_n r)\,\textrm{e}^{-\lambda_n^2 Dt}\,,
  \label{eq:a.15}
\end{equation}
and the $A_n$ are again determined by the initial condition,
corresponding to an expansion of $\rho(r,0)$ into a
Fourier-Bessel series,
\begin{equation}
  A_n = \frac{\int_0^c \textrm{d}r\ r\ J_0(\lambda_n r)\rho(r,0)}
{\int_0^c \textrm{d}r\ r\ J_0^2(\lambda_n r)} = 
\frac{2 J_0(\lambda_n b)}{c^2 J_1^2(\lambda_n c)}\,.
  \label{eq:a.16}
\end{equation}
This yields
\begin{equation}
  p(\mathbf{r},t) = \frac{1}{\pi c^2}
\sum_{n=1}^\infty \frac{J_0(\lambda_n r)J_0(\lambda_n
  b)}{J_1^2(\lambda_n c)}\,\textrm{e}^{-\lambda_n^2 Dt}\,.
  \label{eq:a.17}
\end{equation}
Equation (\ref{eq:pxyt}) follows by setting $c=R$, $b=(1-\epsilon)R$
and $\lambda_n = \alpha_n/R$.

%\bibliographystyle{apsrev4-1}
%\bibliography{surfaces}

\begin{thebibliography}{27}
\bibitem [{com()}]{comm:ishift}%
  \BibitemOpen
  \href@noop {} {}\bibinfo {note} {{The offset corresponding to the shift is
  associated with the preset setpoint current that is mainly related to
  tunneling to/from the bare Ag surface.}}\BibitemShut {Stop}%
\end{thebibliography}


%

\end{document}